\documentclass[aps,prl,twocolumn,footinbib,floatfix]{revtex4-2}

\usepackage{CJK}
\usepackage{amsmath}
\usepackage{amsfonts}
\usepackage{bm}
\usepackage{soul}
\usepackage[caption=false,subrefformat=parens,labelformat=parens]{subfig}
\usepackage{color} 
\usepackage[table,dvipsnames]{xcolor}
\usepackage[colorlinks,linktocpage,bookmarks=false]{hyperref}
\usepackage{booktabs}

\newcommand\myshade{85}
\definecolor{myrulecolor}{RGB}{150,20,0}
\colorlet{mylinkcolor}{violet}
\colorlet{mycitecolor}{YellowOrange}
\colorlet{myurlcolor}{Aquamarine}
\hypersetup{
	linkcolor  = myurlcolor!\myshade!black,
	citecolor  = mycitecolor!\myshade!black,
	urlcolor   =  myrulecolor!\myshade!black,
}

\usepackage{graphicx}
\graphicspath{{FIG/}}
\usepackage{multirow}
\usepackage{braket}

\newcommand{\beq}{\begin{equation}}
\newcommand{\eeq}{\end{equation}}
\newcommand{\bea}{\begin{eqnarray}}
\newcommand{\eea}{\end{eqnarray}}

\newcommand{\CCO}{{Ca$_{10}$Cr$_7$O$_{28}$}}

\renewcommand\[{\begin{equation}}
\renewcommand\]{\end{equation}}

\newcommand{\bq}{\bm{q}}

\newcommand{\bo}{\bm{0}}

\newcommand{\bB}{\bm{B}}
\newcommand{\bpsi}{\bm{\psi}}

\makeindex
\newcommand{\work}{work}

\newcommand{\thistitle}{Identifying topologically critical band from
	pinch-point singularities in spectroscopy}

\begin{document} 
	\begin{CJK*}{UTF8}{gbsn} 
		\title{\thistitle}

\author{Han Yan (闫寒)}
\email{hy41@rice.edu}
\affiliation{Theory of Quantum Matter Unit, Okinawa Institute of Science and Technology Graduate University, Onna-son, 
	Okinawa 904-0412, Japan}
\affiliation{Department of Physics \& Astronomy, Rice University, Houston, TX 77005, USA}
\affiliation{Smalley-Curl Institute, Rice University, Houston, TX 77005, USA}

		\date{\today}

\begin{abstract}
In this paper, we investigate the relationship between pinch point singularities observed in energy- and momentum-resolved spectroscopy and topologically non-trivial gapless points. We show that these singularities are a universal signature, and that the Berry flux encoded must be $n\pi$ for an $n-$fold pinch point under suitable symmetry protection. Our results apply to most systems and are independent of their microscopic details.
Hence they provide a new way to identify topological phases without requiring detailed knowledge of the microscopic model. Our work can be readily applied in spectroscopy experiments on various platforms.
\end{abstract}

\maketitle
\end{CJK*}


\noindent\textbf{\textit{Introduction. --- }}
Topologically critical gapless points in the electron and magnon band structures are significant and frequently-occurring features of quantum matter. \cite{RevModPhys.82.3045,RevModPhys.83.1057,McClarty2022ARCMP,BansilRevModPhys,PhysRevLett.61.2015,PhysRevLett.95.226801,PhysRevLett.95.146802,PhysRevLett.98.106803,PhysRevB.75.121306,PhysRevB.79.195322,Bernevig2006,PhysRevLett.104.066403,Bernevig2013}.
%
Usually,
to unambiguously 
determine if a gapless point
is accidental or 
topological,
one needs to 
reconstruct the Hamiltonian by
reading
the band dispersion relations
from the energy
and momentum-resolved spectroscopy,
while also acquiring a significant amount of microscopic information including
lattice structure, symmetries \textit{etc}.
Such spectroscopy techniques include 
angle-resolved photoemission spectroscopy (ARPES) \cite{RevModPhys.75.473}
and scanning tunneling microscopy (STM) for electron bands \cite{RevModPhys.59.615,RevModPhys.79.353},
and inelastic neutron scattering (INS) for magnon bands \cite{Squires2012Neutron}, and polariton photoluminescence (PP) for photonic lattices \cite{Mili2019PhysRevX}.

\begin{figure*}[th!]
	\centering
	\includegraphics[width=0.9\textwidth]{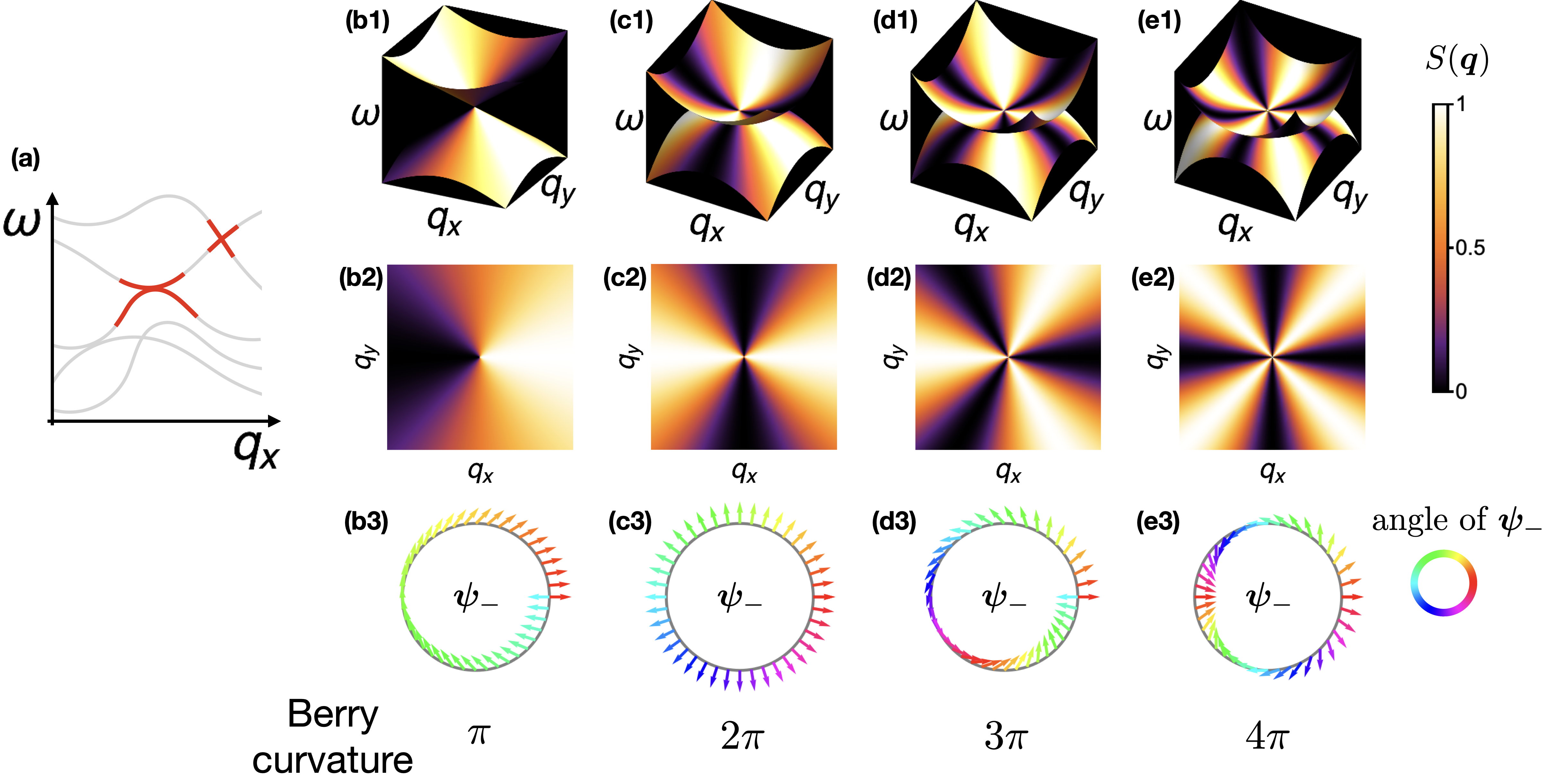}
	\caption{
		Schematic illustration of $n-$fold pinch points.
		(a) Band touching scenarios discussed in this paper, highlighted in red.
		(b-1) 1-fold pinch points imprinted on the two bands.
		(b-2) the spectroscopy density distribution on the lower band.
		(b-3) The corresponding configuration of the wavefunction, which winds $\pi$ around the pinch point. It encodes a Berry curvature of $\pm \pi$ at the gapless point.
		(c,d,e) 2,3,4-fold pinch points illustrated same way as (b).
	} 
	\label{Fig_1234FPP}
\end{figure*}

In this \work, 
we discuss  a different 
way to unambiguously
identify topologically non-trivial gapless points
from spectroscopy alone,
without knowing much else about the system. 
This approach utilizes
the often-ignored information:
the spectroscopic intensity distribution on the 
bands.
Due to the winding of the wave-function around a 
topologically  non-trivial gapless point,
the spectroscopy intensity
on the two bands can show a universal, characteristic singular
pattern which we call an $n-$fold pinch point [Fig.~\ref{Fig_1234FPP}] \cite{PhysRevLett.79.2554}.
Further more, if 
the system admits a suitable symmetry,
the pinch point is 
 \textit{guaranteed} to be topologically critical
and encodes a Berry curvature of $n\pi$.

\begin{table}[ht!] 	
	\caption{
		A   survey of known  experiments exhibiting the pinch points  [Figs.~\ref{Fig_1234FPP},~\ref{Fig_pinch_change}]. 
		For items labeled with a star ($\star$), the pinch points can be seen directly from or inferred from the  references.
		For items without the star, there are no direct observation of the pinch points due to technological limits. But we predict that the pinch points can be observed in principle.
		The gapped pinch point case means the two bands have a small gap opening as shown in Fig.~\ref{Fig_pinch_change}(c).
		``2-fold splits'' means the 2-fold pinch point splits into two copies of 1-fold pinch point, as shown in Fig.~\ref{Fig_pinch_change}(b).
		See first paragraph of main text for abbreviations for experimental methods. 
		\label{Table_materials}}
	\begin{tabular}{rcccc} \toprule
		&Pinch point & Material & Experiment & Ref.  \\ \midrule
		$\star$	&1-fold & graphene & ARPES& \cite{Bostwick2006NatPhys,Fedorov2014,Ludbrook2015,Cheng2015} \\[0.3em]
		$\star$	&1-fold & \begin{tabular}{@{}c@{}}graphene \\ (quasicrystal)\end{tabular}   & ARPES & \cite{Ahn2018Science} \\[0.6em]
		$\star$	&1-fold &  CoTiO$_3$ & INS & \cite{Yuan2020PhysRevX,Elliot2021NatCumm} \\[0.3em]
		$\star$	& 1-fold, gapped &  YMn$_6$Sn$_6$   &  INS & \cite{Zhang2020PhysRevB} \\	[0.3em]
		$\star$	& 1-fold, gapped & CrI$_3$   &  INS & \cite{Chen2018PhysRevX,Chen2021PhysRevX} \\	[0.3em]		
		$\star$	& 1-fold, gapped & CrBr$_3$   &  INS & \cite{Cai2021PhysRevB} \\	[0.3em]		
		$\star$ & 1-fold, gapped &CrSeTe$_3$ & INS & \cite{Zhu2021SciAdv}  \\	[0.3em]
		$\star$ & 1-fold, gapped &Fe$_3$Sn$_2$& ARPES & \cite{Ye2018Nature}  \\	[0.3em]
		$\star$	&2-fold &  \begin{tabular}{@{}c@{}}graphene \\ (bilayer)\end{tabular}   & STM & \cite{JouckenPhysRevB2020} \\[0.6em]
		$\star$&2-fold &Nd$_2$Zr$_2$O$_7$   & INS & \cite{Petit2016NatPhys,Lhotel2018NatComm,XuLakePhysRevLett2020,XuLakePhysRevB2019} 		\\ [0.3em]
		&2-fold &\CCO & INS & \cite{Balz2016,Balz2017-PRB95,Balz2017-JPCM29}\\[0.3em]
		$\star$&2-fold, gapped & FeSe & ARPES & \cite{Liu2012,Wang2016NatMatt}   \\ [0.3em]
		$\star$&2-fold, gapped & CoSn & ARPES, STM & \cite{Kang2020,Liu2020NatCom} \\[0.3em]
		$\star$&2-fold, gapped &Lu$_2$V$_2$O$_7$   & INS & \cite{Mena2014PRL} 		\\ [0.3em]
		& 2-fold, gapped & Cu(1,3-bdc) &  INS & \cite{Chisnell2015PhysRevLett} \\[0.3em]
		$\star$ &  \begin{tabular}{@{}c@{}}2-fold  splits  \\  or opens gap \end{tabular}   &   \begin{tabular}{@{}c@{}} photonic orbital  \\  graphene \end{tabular}  & PP  &\cite{Mili2019PhysRevX}\\
		\bottomrule
	\end{tabular} 
\end{table}

Experimentalists have actually 
observed this universal pattern in a various materials [Table.~\ref{Table_materials}],
although the hidden connection has not been
discussed much.
The simplest case,  $1-$fold pinch point,
is actually a Dirac cone,
and has been shown in ARPES experiments \cite{Bostwick2006NatPhys,Fedorov2014,Ludbrook2015,Ahn2018Science} on
several  graphene-based materials \cite{PhysRevLett.95.226801,Novoselov2005,Zhang2005}.
The  $2-$fold pinch points
appear in bilayer graphene  \cite{JouckenPhysRevB2020},
FeSe \cite{Liu2012,Wang2016NatMatt}, and various frustrated lattice materials
\cite{Petit2016NatPhys,Lhotel2018NatComm,XuLakePhysRevLett2020,XuLakePhysRevB2019,Balz2016,Balz2017-PRB95,Balz2017-JPCM29,Benton2016PhysRevB,Kshetrimayum2020,Sonnenschein2019,pohle2021arxiv,Chisnell2015PhysRevLett,Kang2020} . 
The  $n-$fold pinch point's implication of the underlying Gauss's law
has been a focus in classical and quantum spin liquids   \cite{moessner98-PRB58,huse03,henley10,Benton2016,PremPhysRevB2018,Benton2016PhysRevB,HY2018PhysRevB,PhysRevLett.124.127203,yan2022phonon,PhysRevResearch.4.023175,PhysRevLett.127.107202},
but the connection to Berry curvature
has been rarely mentioned.

The   universal patterns of pinch points
 has a high application value.
Its advantage relies  on the fact that 
it does not require one to know much about the 
microscopic details of the matter or to
reconstruct the full Hamiltonian.
Also, this is particularly useful
for two touching-bands with quadratic or higher-order gap-opening
dispersion, where the dispersion  alone is not a
very distinguishing factor like the Dirac cones.
Another application is to twisted bilayer systems \cite{Cao2018,Cao2018nature2,Lisi2020}, 
where the full Hamiltonian for all bands is practically impossible to reconstruct.

\mbox{}\\ \textbf{\textit{Dirac cone as 1-fold pinch point. --- }}
We start by introducing the ``pinch point''.
In this \work we work on 2D systems,
but its generalization to 3D is fairly straightforward.
The simplest case ---  $1-$fold pinch point ---
is imprinted on the most common ingredient 
of topological band systems: the Weyl/Dirac cone \cite{shivam2017neutron,Elliot2021NatCumm,McClarty2022ARCMP}.
 
Consider a local region in the reciprocal momentum space, 
where  two bands have a degenerate point set at $\bm{q} = \bm{0}$ [cf. Fig.~\ref{Fig_1234FPP}(a)].
In the neighborhood of $\bm{q} = \bm{0}$, they are also gapped from other bands,
so we can    focus on the two-band subsystem only.

Spectroscopy measures the band structure
as well as the intensity distribution of certain correlation function on each band.
We denote the  upper and lower  band's dispersion relations as  $\omega_-(\bm{q})$ and  $\omega_+(\bm{q})$.
The energy and momentum-resolved spectroscopy 
of the two bands are (assuming infinitely fine resolution) 
\[
\begin{split}
	\mathcal{S}_+(\omega,\bq) = \delta(\omega -\omega_+ (\bq) ) S_+(\bq), \\
	\mathcal{S}_-(\omega,\bq) = \delta(\omega -\omega_- (\bq) ) S_-(\bq).
\end{split}
\]
Here, we separated the dispersion  $\delta(\omega -\omega_\pm (\bq) )$
and the intensity distribution $S_\pm(\bq)$ for future convenience.
The intensity $S_{\pm}$  usually measures the amplitude of the wavefunction
in a particular  basis.
For example, in case of graphene, it measures 
$\braket{(c_A^\dagger + c_B^\dagger)(c_A + c_B)}$ of the band wavefunction, where $A,B$ are the two sublattice indices, so the basis 
measured is $c_A + c_B$.
The wavefunction $(1/\sqrt{2},1/\sqrt{2})^T$ has maximal intensity, while $(1/\sqrt{2},-1/\sqrt{2})^T$ has zero.

The $1-$fold pinch point
refers to the spectroscopic intensity $S_\pm(\bq)$
distribution on the Dirac cone as illustrated in Fig.~\ref{Fig_1234FPP}(b2).
The intensity distribution only depends on the angle around 
the gapless point.
On one band, it reaches  zero on one side, and maximum on the other.
The intensity varies smoothly except at the gapless point, 
where it becomes singular (i.e. not continuous).
The other band 
has a similar pattern of  
intensity distribution, but
the strong and weak regions switch sides.
This pattern has in fact been observed in various experiments,
as we summarized in Table.~\ref{Table_materials}.

\mbox{}\\ \textbf{\textit{$n-$fold pinch point. --- }}
The  pattern of $1-$fold pinch point
can  be generalized to $n-$fold pinch point.
The cases of $n=1,2,3,4$ are illustrated in Fig.~\ref{Fig_1234FPP}.
The upper row panels show the entire energy and momentum resolved spectroscopy, and the mid row panels show  the  intensity distribution $S_\pm(\bq)$ without the dispersion.

The crucial ingredient of the $n-$fold pinch point
is the  singularity 
 in  the  intensity distribution $S_\pm(\bq)$.
Near $\bm{q} = \bm{0}$, the  intensity 
only depends on the angle $\theta$ around $\bm{0}$. 
An $n-$fold pinch point has $n$ dark wings where the intensity is low,
and $n$ bright wings where the intensity is high.
The most symmetric form of the intensity distribution,
by choosing a suitable angle as $\theta=0$, is
\[
\label{eqn_correlation_definition_n_general_first}
\begin{split}
S_+(\bq) =  A\cos^2(n\theta/2),  \\
S_-(\bq) =    A\sin^2(n\theta/2)  .
\end{split}
\]
Here, $A$ is just a scalar signifying the overall intensity.
$S_\pm(\bq)$  at $\bq = \bm{0}$ is singular,
since one obtains different values of $S_\pm(\bm{0})$ by approaching it from different directions.

For  a specific lattice model, the intensity distribution can be 
mildly distorted up to an isomorphic mapping.
The more general form is
\[
\label{eqn_general_pp}
\begin{split}
S_+(\bq)  = A\cos^2(n \Theta(\theta)/2),\\
 S_-(\bq)=A \sin^2(n \Theta(\theta)/2),
 \end{split}
\]
where $\Theta(\theta)$ is a
smooth, monotonically increasing function 
as a bijection from $[0, 2\pi)$ to itself.
\[
\label{eqn_f_condition}
\Theta(0)=0, \; \Theta(2\pi) = 2\pi ,\Theta'(\theta)>0	 .
\]

\begin{figure}[th!]
	\centering
	\includegraphics[width=0.95\columnwidth]{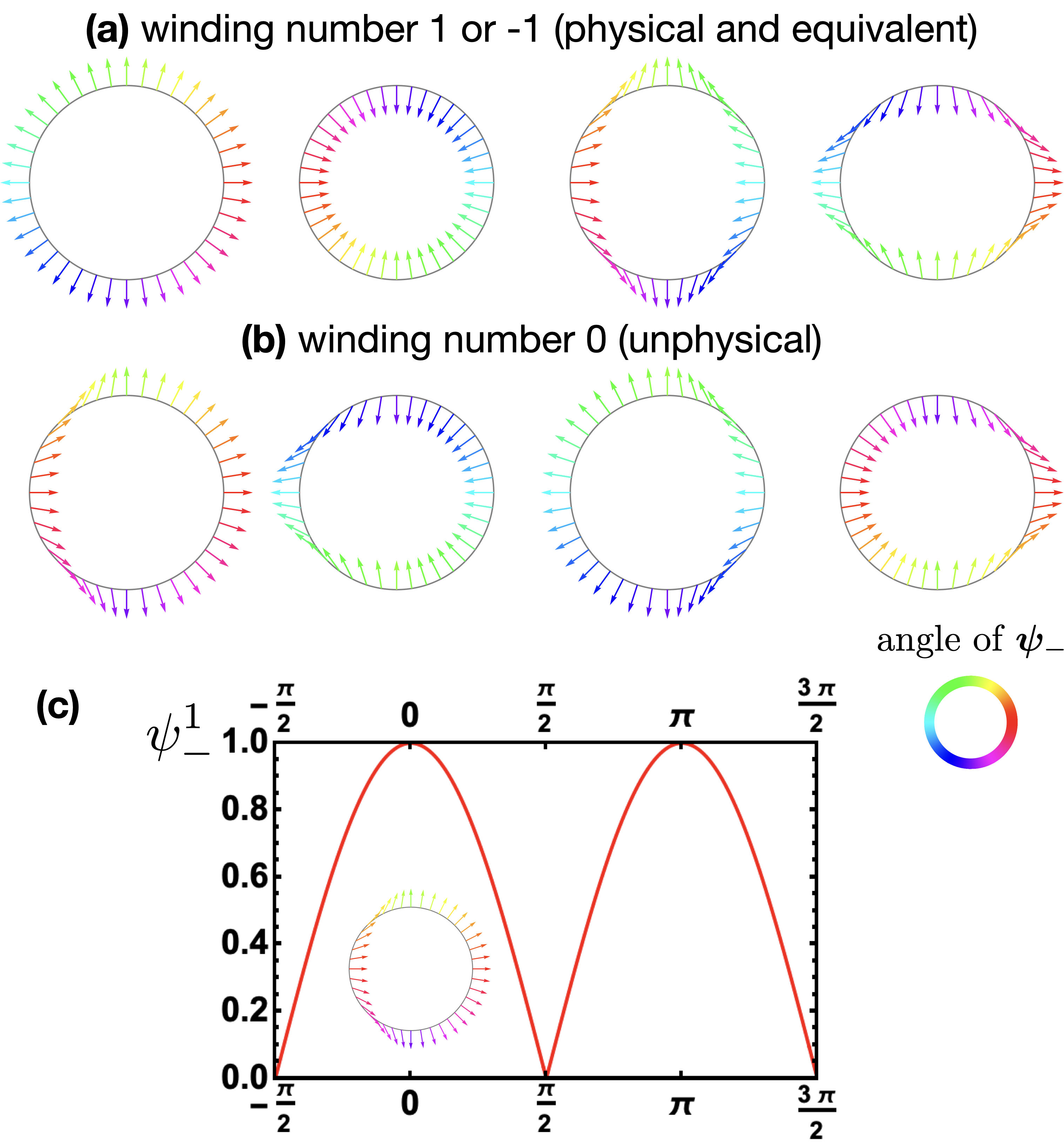}
	\caption{
	Representative continuous winding configurations of $\bpsi_-$ consistent with a 2-fold pinch point.
	(a) $\bpsi_-$ configurations with winding number $\pm1$, which are also smoothly varying.
	(b) $\bpsi_-$ configurations with winding number $0$, which are not smoothly varying.
	(c) Plot of $\psi_-^1$ as a function of angle $\theta$, showing non-smoothness at $\theta = \pi/2,\ 3\pi/2$.
		\label{Fig_all_winding_config}
	}
\end{figure}

The main message of this paper is   that
such gapless $n-$fold pinch point
is guaranteed to be topologically nontrivial.
By ``non-trivial'' we mean
the degenerate point is not accidental, and 
generally carries a non-zero Berry curvature,
except for some extremely fine-tuned cases.
A much stronger result is that if  the system 
additionally admits a  suitable
symmetry  (for example, time reversal symmetry for spinless electron  systems),
then the  Berry flux encoded must be  $n\pi$.

A   remark is in order before we proceed to the proof.
In this work we assume that there is no ``extrinsic''
factors from the coupling between the system and the 
probing particles (photons or neutrons), 
which may also exhibit pinch point patterns.
For ARPES, if the electron is in an anisotropic orbit, 
 the photon-electron coupling will pick up an angle-dependent projector depending on the polarization of photons, which may yield pinch point patterns.
A very detailed discussion can be found in Ref.~\cite{Moser2017}.
One needs to choose the photon polarization (or use unpolarized photons) properly to avoid such effects.
The same principle applies to INS --
for example, polarized neutrons couple to spin-1/2  dimers in a fixed direction can also have similar projectors.
However, these extrinsic, ``fake'' pinch points are often not present in actual experiment, or at least avoidable in principle. 
In all examples given in Table.~\ref{Table_materials}, one does not need to worry about them.

\mbox{}\\ \textbf{\textit{Pinch points are topologically critical. --- }}
We now prove that under suitable symmetries, 
the $2-$fold pinch-point singularity
is  associated with a gapless point with
Berry flux $\pm 2 \pi$ encoded . 
Here we take the symmetry be  time reversal   ($\mathcal{T}$) for the spinless electrons,
which  forces the Hamiltonian to be real.
The proof can be easily generalized to  $n-$fold pinch-points and other proper symmetries (see end of section).

The physical picture is the following.
The $n-$fold pinch point pattern (Fig.~\ref{Fig_1234FPP}(b1-e1)) indicates that the spectroscopy 
intensity follows a squared-sine function.
This requires that the wavefunction of the corresponding band, which is a two component real vector,  
rotates in a sine/cosine manner on a  loop  around the gapless point,
and accumulate a  quantized total rotation angle $n\pi$ (Fig.~\ref{Fig_1234FPP}(b3-e3)) at the end.
This is exactly the origin of the Berry flux.

Now, onto the formal proof,
we first set up the model
with a few simplifications 
without  affecting its topological features.
%
In the vicinity of the pinch-point,
we consider the relevant subsystem with two   degrees of freedom
$ \psi^1 ,\  \psi^2$,
and their corresponding two-level Hamiltonian.
The upper and lower bands
correspond to two wavefunctions
written as two orthonormal, unit vectors of complex entries
\[
\label{eqn_wave_fn} 
  \bm{\psi}_+(\bq ) =
\begin{pmatrix}
\psi_+^1\\ \psi_+^2
\end{pmatrix}, \qquad
  \bm{\psi}_-(\bq ) =
\begin{pmatrix}
\psi_-^1 \\ \psi_-^2
\end{pmatrix} . 
\]
The system is then described by a   Hermitian Hamiltonian
\[
\label{eqn_Hamiltonian_abstract}
\mathcal{H}(\bq) = \bm{S}
\begin{pmatrix}
\omega_+(\bq) & 0\\
0 & \omega_-(\bq) 
\end{pmatrix}
\bm{S}^\dagger.
\]
where $\bm{S}= \left(  \bpsi_+, \bpsi_- \right)$ is the eigenvector matrix,
and $\omega_\pm(\bq)$ are the two bands' dispersion relations.

Under the $\mathcal{T}$ symmetry,
the eigenvectors and Hamiltonian are real.
Because  the spectroscopy 
intensity follows a squared-sine function,
it requires 
\[
\label{eqn_psi_1_mt}
|\psi_-^1(\bq)| \propto |q^x|/q = |\sin\theta|	.
\]
This puts strong constraint on the possible eigenvector configurations. Some of them are listed in Fig.~\ref{Fig_all_winding_config}(a,b).

We also require 
 the Hamiltonian to be \textit{smooth},
i.e., \textit{continuous} and \textit{differentiable}  to any order.  
This is generically true for systems with short-rang interactions,
and not so only for systems with certain   fine-tuned long-range interactions.
Hence the requirement applies to
 most physically realistic systems \cite{Bloch1929,BouckaertPhysRev.50.58}.
It   plays a key role 
in eliminating the physically unrealistic cases shown in Fig.~\ref{Fig_all_winding_config}(b),
because in those cases, the eigenvector has to go though points where unsmooth change of its component(s) is bounded to happen, even though there is no gap closing at those points. 
One of such unsmooth components is illustrated in Fig.~\ref{Fig_all_winding_config}(c).
 A more detailed, technical analysis is presented in the appendix. 

Therefore, 
for $\psi_-^1$ to vary smoothly, 
it has to take the form
$\psi_-^1 = \sin\theta$
over the entire circle, up to an overall minus sign.
After a similar analysis on $\psi_-^2$ and $\bpsi_+$,
we can conclude that 
the viable eigenvectors in $\bq -$space are those in  
Fig.~\ref{Fig_all_winding_config}(a), which are all topologically equivalent in eyes of Berry flux at the gapless point.
We may pick the eigenvectors to be
\[
\label{EQN_cos_sin_wavefunction}
\begin{split}
	& \bpsi_+(\bq) =(- \cos\theta, \sin\theta )=\frac{1}{q} (-q^y,q^x), \\
	& \bpsi_-(\bq) = (\sin\theta, \cos\theta  )=\frac{1}{q} (q^x,q^y). \\
\end{split}
\]

The eigenvectors and eigenvalues completely determines the Hamiltonian (cf. Eq~\eqref{eqn_Hamiltonian_abstract}).
Written as 
 an effective magnetic field coupled to Pauli matrices, it is
\[
\label{eqn_2fpp_ham_B_pauli}
\begin{split}
\mathcal{H}(\bq) =  
&\frac{\omega_+(\bq)+\omega_-(\bq)}{2} \mathbb{I} \\
&+ \Delta(\bq)  \frac{(q^{x})^2-(q^{y})^2}{2q^2} \sigma_z
- \Delta (\bq)  \frac{q^x q^y}{q^2} \sigma_x		\\[1em]
\equiv &\frac{\omega_+(\bq)+\omega_-(\bq)}{2} \mathbb{I} + \bm{\sigma}\cdot\bm{B}(\bq),
\end{split}
\]
where 
$\Delta(\bq) = \omega_+(\bq) - \omega_-(\bq)$
is  the energy gap.

From the Hamiltonian we can read off the effective magnetic field $\bm{B}(\bq)$ to be
\[
\label{eqn_2fpp_B}
\bm{B}(\bq) =\Delta(\bq) \left(-\frac{q^x q^y}{q^2} ,0,  \frac{(q^{x})^2-(q^{y})^2}{2q^2} \right)	.
\]

\begin{figure}[t]
	\centering
	\includegraphics[width=0.95\columnwidth]{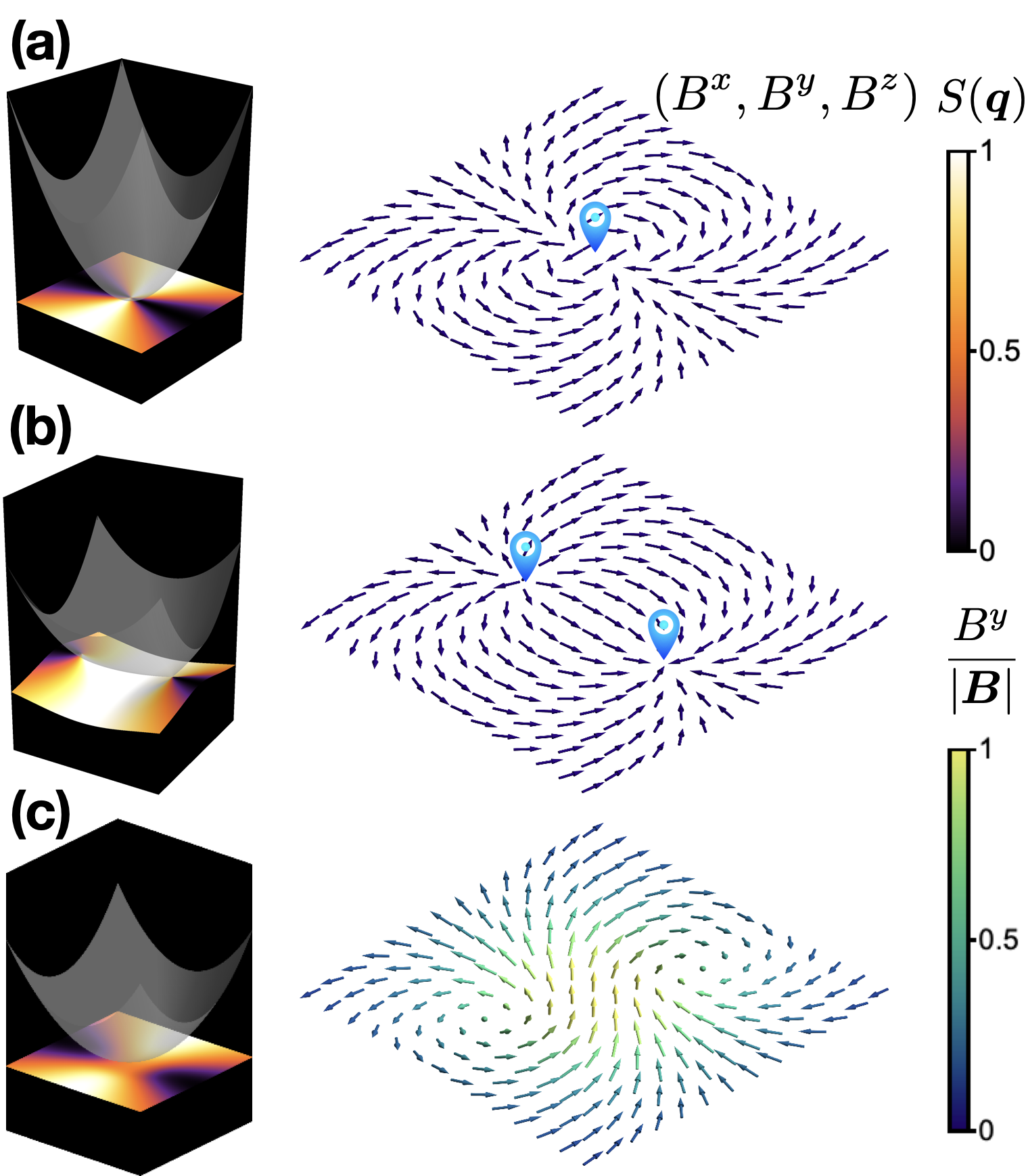}
	\caption{Transformation of 2-fold pinch point  and corresponding effective magnetic field configuration.
		For better visibility we set the upper band to be transparent and focus on the lower band.
		(a) The 2-fold pinch point and winding number-2 magnetic field in the effective Hamiltonian.
		(b) Upon introducing $\delta B^z$, the gapless point splits into two Dirac points. The 2-fold pinch point splits into two 1-fold pinch point.
        (c) Upon introducing $\delta B^y$, the two bands become gapped and pick up a Chern number.
		The singularity disappears.
	} 
	\label{Fig_pinch_change}
\end{figure}

The crucial property is that,
on a loop around the pinch point,
the two components $(B^x,B^z)$  as a 2D vector field form a vortex of winding number 2.
In Fig.~\ref{Fig_pinch_change}(a),
the normalized $(B^x,B^z, B^y)/B(\bq)$ is plotted.
Here we swapped   $B^y$ and $B^z$ for better visualization.
This is known to encode a Berry curvature of $\pm 2 \pi$ \cite{Bernevig2013}.
Another way to see this is to directly compute the  Berry flux enclosed by a loop  $\bq=q(\cos\theta, \sin\theta)$ around the gapless point, defined as 
\[\label{EQN_berry_flux_def}
C_\pm = \int_0^{2\pi} \text{d}\theta\ i  \bm{\psi}_\pm^\dagger \cdot \partial_\theta \bm{\psi}_\pm .
\] 
We  conclude our proof here.

This proof can be intuitively generalized to a general $n-$fold pinch point. 
In Fig.~\ref{Fig_1234FPP}, we plot the smoothly varying $\bm{\psi}_-$ for different cases. Note that for odd $n$, $\bm{\psi}_-$ needs anti-periodic boundary condition instead.

Without the symmetry protection,
the different components of an 
eigenvectors can have different
complex phases, 
so the proof above does not apply any longer,
and
the Berry-flux at the gapless point is generally
not quantized. 
One example of such scenarios 
is to add  different phases to $\psi^1$ and $\psi^2$ in Eq.~\eqref{EQN_cos_sin_wavefunction},
which can yield a finite Berry flux contribution when plugged into Eq.~\eqref{EQN_berry_flux_def},
or even render to total Berry zero.
In this case, $\bm{B}(\bq)/B$
still travels back and forth twice from the north pole to south pole on the unit sphere  for a path of $\bm{q}$ around the pinch point. 
It can be, however,  not two great circles, but some general curve.
The solid angle enclosed by the path, which is the Berry flux, is then not quantized.

Finally, the picture above also shows that
other symmetry protections  can also work, 
if the normalized effective magnetic field is restricted to move on a fixed great circle between the two poles.

\mbox{}\\ \textbf{\textit{Splitting and gapping the pinch point. --- }}
How the topologically critical gapless points 
transform under different perturbations
is a well-studied topic.
In this section we revisit some of these transformations,
with a focus on the corresponding pinch point phenomenology.
 
%
Fig.~\ref{Fig_pinch_change}(b) shows
a 2-fold pinch point
spliting into two 1-fold pinch points (Dirac cones).
Correspondingly,
the Berry curvature of $2\pi$ 
is also split into the two 1-fold pinch points each carrying Barry flux $\pi$,
assuming the proper symmetry protection.
In this process the overall Berry flux is conserved.

Using  our model for demonstration,
this can be done  by  introducing  a  perturbation  of constant $B^z> 0$ 
for the Hamiltonian in \eqref{eqn_2fpp_ham_B_pauli}, and take $\Delta(\bq) = cq^2$,
\[
\bB_\text{z-tuned}(\bq) = \left(-c q^x q^y ,0,  \frac{c}{2} ((q^{x})^2-(q^{y})^2)+ \delta B^z \right)	.  
\] 

The original gapless point at $\bq = \bo$ is then split into two 
 linearly dispersive gapless points
at $\bq = (0, \pm \sqrt{2\delta B^z / c})$.
%
 the original winding number-$2$ vortex 
 of $\bB(\bq)$ 
 splits into two vortices, each with winding number $1$ [Fig.~\ref{Fig_pinch_change}(b)], consistent with the Berry flux conservation. 
  
The critical gapless point can also be gapped,
and  induces well-defined, opposite  non-zero Berry curvature on 
the two bands locally.
We can consider perturbing the effective magnetic field in the following way,
\[
\bB_\text{y-tuned}(\bq) = \left(-c q^x q^y ,\delta B^y,  \frac{c}{2} ((q^{x})^2-(q^{y})^2)  \right)	. 
\]
As a consequence, the normalized $\bB_\text{y-tuned}/B_\text{y-tuned}$
form half a skyrmion, or a meron, as illustrated in Fig.~\ref{Fig_pinch_change}(c).
Since the skyrmion is of winding number $2$,
the two bands get local
Berry curvature $\pm 2 \pi$.
The $\pm$ sign depends on the sign of $\delta B^y$,
and cannot be distinguished from the spectroscopy pattern.

The pinch point singularity disappears as the gap opens.
The spectroscopic intensity on two bands
becomes smooth at the center, 
but gradually recovers the pinch point pattern 
when zoomed out. 
An example of this  will be discussed in detail in a separate work studying the  Kagome model \cite{yan2023pinch}. Similar examples can also be found in Refs.~\cite{PhysRevLett.103.046811,PhysRevB.78.245122}.

\mbox{}\\ \textbf{\textit{Discussion. --- }}
The main message of this work is that 
the $n-$fold pinch points observed in energy-momentum resolved spectroscopy 
indicate that the gapless point is  topologically critical,
and encodes Berry flux $n\pi$
if there is a suitable symmetry protection.
We have proven this for time reversal symmetry, and also provided a  survey of experiments (Table.~\ref{Table_materials}) that observe  the universal phenomenon. 
This result does not rely on further microscopic details 
of the system, hence has great potential in experimental application across several platforms including ARPES, STM, and INS.

Another useful lesson  is  that the reversed conclusion is \textit{often but not always} true: a topologically critical gapless point \textit{may} appear as a $n-$fold pinch point. 
Some of these cases has been discussed in Refs.~\cite{Elliot2021NatCumm,Mucha2008PhysRevB,shivam2017neutron},
although the connection to pinch points were not mentioned.
It is not always true, because  the winding of certain wavefunction  may not be captured by the correlation function measured by spectroscopy.  
The Honeycomb and Kagome lattice magnon models studied in Ref.~\cite{PhysRevB.93.014418,PhysRevB.92.144415,HY2018PhysRevB} are such   examples, in which some critical gapless points appear as 1FPP and 2FPP, but some are completely dark in the structure factor due to extrinsic, accidental cancellation of form factor.

The bigger picture we gain from this work is that 
the spectroscopy contains  a huge amount of information,
of which a lot are still waiting to be exploited.
For example,
instead of points,
other topologically critical  loci  of band degeneracy should also 
manifest universal, characteristic patterns.
This idea also applies to interacting or non-Hermitian Hamiltonians.
Developing the zoology of them will be an important and useful piece of phenomenological study.

\section*{Acknowledgment}
We specially thank Nic Shannon for inspiring discussions 
at the beginning stage of this work,
and Owen Benton for  enlightening discussions
on Berry flux.
We also  thank   Andreas Thomasen 
for his helpful review of the paper.
H.Y. is supported by the Theory of Quantum Matter Unit at Okinawa Institute of Science 
and Technology, the Japan Society for the Promotion of Science (JSPS) Research Fellowships 
for Young Scientists, and the
National
Science Foundation Division of Materials Research under
the Award DMR-191751  at 
different stages of this project.

\bibliography{reference}


\clearpage
\setcounter{equation}{0}
\setcounter{figure}{0}
\setcounter{table}{0}
\makeatletter
\renewcommand{\theequation}{S\arabic{equation}}
\renewcommand{\thefigure}{S\arabic{figure}}
\renewcommand{\bibnumfmt}[1]{[#1]}
\renewcommand{\citenumfont}[1]{#1}

\onecolumngrid

\begin{center}
	\large{\textbf{Appendix for ``\thistitle"}}
\end{center}

In this section, we give a detailed proof of Eq.~\eqref{EQN_cos_sin_wavefunction}.
That is, the squared-sin pattern of the spectra distribution indicates that the corresponding eigenvector has to be equivalent to Eq.~\eqref{EQN_cos_sin_wavefunction}, or those Fig.~\ref{Fig_all_winding_config}(a). 
The configurations in Fig.~\ref{Fig_all_winding_config}(b) are forbidden.
 
To start with, we have the freedom to
choose the basis $(1,0)$ to be the one
on which the 
spectroscopy measures the amplitude of the wavefunction. 
This is because the spectroscopy is measuring the expectation value of certain operator of scalar degree of freedom.
So we have
\[
\label{eqn_correlation_definition_S}
\begin{split}
	S_+(\bq) \propto | (1,0)\cdot(\psi^{1 }_+,\psi^{2 }_+)^\mathrm{T}(\bq) |^2 =  |\psi_+^1(\bq)|^2 ,\\
	S_-(\bq) \propto |(1,0)\cdot(\psi^{1 }_-,\psi^{2 }_-)^\mathrm{T}(\bq)|^2 = |\psi_-^1(\bq)|^2 .
\end{split}
\]
In the example of graphene, $\psi^1  \sim  c_A + c_B$ 
is measured by ARPES,  and $\psi^2  \sim  c_A - c_B$.
%
%
We also assume that the pinch point pattern is in its most symmetric form 
\[
\label{eqn_correlation_pinch_S}
\begin{split}
	S_+(\bq)= A\cos^2(\theta)  =A \frac{(q^y)^2}{q^2} ,\\
	S_-(\bq) =  A\sin^2(\theta) =A \frac{(q^x)^2}{q^2}	 .
\end{split}
\]
A small distortion of the pinch point 
in form of Eq.~\eqref{eqn_general_pp}
does not affect  our conclusions on its  topological features.




Combining Eq.~\eqref{eqn_correlation_definition_S} and Eq.~\eqref{eqn_correlation_pinch_S},
we see that 
\[
\label{eqn_psi_1}
|\psi_-^1(\bq)| \propto |q^x|/q = |\sin\theta|	.
\]
Since $\bm{\psi}_-$ is a unit vector, we then also know
\[
\label{eqn_psi_2}
|\psi_-^2(\bq)|  \propto |q^y|/q=  |\cos\theta|	,
\]
There  are several ways to arrange 
\textit{continuously} varying $\bm{\psi_-}$ on a circle  around the pinch point 
and satisfy conditions of Eq.~\eqref{eqn_psi_1} and Eq.~\eqref{eqn_psi_2}.
Some representatives are illustrated in Fig.~\ref{Fig_all_winding_config}(a,b).

The configurations in the same row are   physically identical,
since they can be
related to each other by introducing minus signs to the basis $\psi^1$ or $\psi^2$.
The configurations in  Fig.~\ref{Fig_all_winding_config}(a), compared to 
those in Fig.~\ref{Fig_all_winding_config}(b),
are physically different:
the first row has winding number $\pm 1$,
and the second  has  winding number $0$.
%
%
%
%

A key observation is   that  $\psi_-^1$ configurations with 
zero winding number are not smooth, hence
do not yield a \textit{smooth} Hamiltonian.
%
%
Taking the first case of Fig.~\ref{Fig_all_winding_config}(b)
as an example,
$\psi_-^1$ takes a sharp turn at $\theta = \pm \pi/2$
as shown in the Fig.~\ref{Fig_all_winding_config}(c).
However, $\psi_-^2=\pm1$, and   varies smoothly following a sine curve at these two point.
The Hamiltonian is 
\[
\label{eqn_Hamiltonian_abstract_S}
\mathcal{H}(\bq) = \bm{S}
\begin{pmatrix}
	\omega_+(\bq) & 0\\
	0 & \omega_-(\bq) 
\end{pmatrix}
\bm{S}^\dagger.
\]
where $\bm{S}= \left(  \bpsi_+, \bpsi_- \right)$.
So its off-diagonal terms component,
containing $\omega_-\psi_-^1 \psi_-^2$,
is then not smooth.
%

One may want to amend this problem by 
``softening'' the sharp turn of $\psi_-^1$  at $\theta  = \pi/2,\ 3\pi/2$,
hoping that it will   yield  a distorted pinch point described by some  function $\Theta(\theta)$ in Eq.~\eqref{eqn_general_pp} and has a smooth Hamiltonian.
However, in this case we always have 
$\psi_-^{1\prime}=\psi_-^{2\prime}=0$, so that
$\Theta'(\theta_0) = 0$
at the points $\Theta(\theta_0) =  \pi/2,\ 3\pi/2$, 
which is not the proper pinch point pattern   defined in
Eq.~\eqref{eqn_f_condition} anymore.
Such unauthentic 
pinch points can be distinguished in experiments
since  $\Theta(\theta)$ can be measured.

Therefore, 
the eigenvector configurations in Fig.~\ref{Fig_all_winding_config}(b) are forbidden.
It has to be equivalent to Eq.~\eqref{EQN_cos_sin_wavefunction}, or those Fig.~\ref{Fig_all_winding_config}(a),
which encodes a Berry flux at the gapless point as shown in the main text.

\end{document}